\documentclass[aps,twocolumn,amssymb,prl,showpacs,10pt]{revtex4-2}

\usepackage[utf8]{inputenc}
\usepackage{tabularx}
\usepackage{multirow}
\usepackage{url}
\usepackage{amsmath}
\usepackage{amssymb,amsmath}	
\usepackage{gensymb}
\usepackage{graphicx}
\usepackage{mathrsfs}
\usepackage{color}
\usepackage[normalem]{ulem}
\usepackage{float}
\usepackage{lmodern}
\usepackage{soul}
\usepackage[breaklinks,colorlinks,urlcolor=blue,citecolor=blue]{hyperref}

\renewcommand{\vec}[1]{\mathbf{#1}}
\newcommand{\be}{\begin{equation}}

\newcommand{\ee}{\end{equation}}
\newcommand{\bi}{\begin{itemize}}
\newcommand{\ei}{\end{itemize}}
\newcommand{\figg}[1]{Fig.~\ref{fig:#1}}

\newcommand{\ls}{}
\newcommand{\kint}{k_{\rm int}} 
\newcommand{\kintde}{k_{\rm int}d_e} 
\newcommand{\tl}{t_{\rm L}} 
\newcommand{\ttl}{t/t_{\rm L}} 
\newcommand{\de}{d_{e}} 
\newcommand{\epsM}{\epsilon_{\rm M}} 
\newcommand{\epsK}{\epsilon_{\rm K}}
\newcommand{\kjdotb}{k_{\rm J \cdot B}}
\newcommand{\kjcrossb}{k_{\rm J \times B}}

\begin{document}

\title{
Generation of near-equipartition magnetic fields in turbulent collisionless 
plasmas}

\author{Lorenzo Sironi}
\email{lsironi@astro.columbia.edu}
\author{Luca Comisso}
\author{Ryan Golant}
\affiliation{Department of Astronomy and Columbia Astrophysics Laboratory, Columbia University, New York, NY 10027, USA}
\date{\today}

\begin{abstract}
The mechanisms that generate ``seed'' magnetic fields in our Universe and that amplify them throughout cosmic time remain poorly understood. By means of fully-kinetic particle-in-cell simulations of turbulent, initially unmagnetized plasmas, we study the genesis of magnetic fields via the Weibel instability and  follow their dynamo growth up to near-equipartition levels. In the kinematic stage of the dynamo, we find that the rms magnetic field {strength} grows exponentially with rate $\gamma_B \simeq 0.4\,u_{\rm rms}/L$, where $L/2 \pi$ is the driving scale and $u_{\rm rms}$ is the rms turbulent velocity. In the saturated stage, the magnetic field energy reaches about half of the turbulent kinetic energy. Here, magnetic field growth is balanced by dissipation via reconnection, as revealed by the appearance of plasmoid chains. At saturation, the integral-scale wavenumber of the magnetic  spectrum approaches $k_{\rm int}\simeq 12\pi/L$. Our results show that turbulence---induced by, e.g., the gravitational build-up of galaxies and galaxy clusters---can magnetize collisionless plasmas with large-scale near-equipartition fields.   
\end{abstract}

\maketitle
Magnetic fields are everywhere, shaping the Universe on all scales. They pervade even the most 
rarefied spaces in our cosmos \cite{Widrow02,beck_15}, i.e., cosmic voids and filaments, and the intracluster medium (ICM) of galaxy clusters. Yet, the mechanisms that generate magnetic fields in the Universe and govern their amplification throughout cosmic time are still poorly constrained by observations \cite{GR01,kulsrud_08,durrer_13,Subram16}.  

It is generally accepted that weak ``seed'' magnetic fields can be amplified by dynamo processes \cite{BS05,scheko_04,rincon_19,scheko_22,Tobias21}, yet 
the mechanisms that generate these seed fields in initially unmagnetized plasmas are poorly understood. In collisionless unmagnetized plasmas, magnetic fields can be generated from scratch by the Weibel instability \cite{weibel_59,Davidson72,Medvedev99,Silva03,Medvedev06,Silva21,califano_98,califano_06,bret_08}, which taps into the free energy of temperature anisotropies. Hybrid simulations with kinetic ions and fluid electrons \cite{rincon_16,stonge_18} could not capture the physics of the Weibel instability and needed to start from a prescribed initial seed field. Recent fully-kinetic simulations have shown that the Weibel instability can  grow in collisionless plasmas subject to externally-driven turbulence \cite{pucci_21} or the action of a shear flow \cite{zhou_22}. Still, Weibel-generated fields saturate at small amplitudes and small scales. It is yet to be determined whether, under the continuous action of a turbulent flow, Weibel fields can be dynamo-amplified and reach near-equipartition with the turbulent kinetic energy. 

In this {\it Letter}, we perform 
 the first fully-kinetic particle-in-cell (PIC) simulations capturing both the genesis of magnetic fields via the Weibel instability and their dynamo growth  up to near-equipartition levels in turbulent, initially unmagnetized plasmas.
Our simulations show that in the saturated dynamo stage the magnetic field energy is about half of the turbulent kinetic energy and the integral-scale wavenumber of the magnetic  spectrum approaches $\kint\simeq 12\pi/L$ (where $L/2 \pi$ is the driving scale).
Thus, turbulence
can magnetize collisionless plasmas with near-equipartition fields {on scales much larger than characteristic kinetic scales.} 

{\it Method---}We perform large-scale 3D PIC simulations \cite{yee_66,Boris70,umeda_03} with TRISTAN-MP \citep{buneman_93, spitkovsky_05} in a periodic cube of size $L^3$. We consider a pair plasma, which excludes field generation via Biermann battery \cite{biermann_50}, so the only source of seed fields is the Weibel instability \cite{schoeffler_14,schoeffler_16}. 
We drive turbulence on the box scale using an Ornstein–Uhlenbeck process, adding an external acceleration term to the equations of motion of PIC macro-particles (see Suppl.~Mat. and \cite{zrake_12}). The acceleration term is charge-independent and  meant to mimic the effect of turbulent external gravitational forces; its strength is such that the target rms velocity is $u_{\rm rms}\sim 0.2 \,c$. 
We define the turnover time $t_{\rm L}\equiv(L/2 \pi)/u_{\rm rms}$, and set the decorrelation rate 
of the driver to
 $\gamma_{\rm corr} \simeq  1.5/\tl$. In most of our runs, we drive purely vortical (solenoidal) turbulence, but we also present simulations with different ratios of solenoidal and compressive power, which we quantify by the dimensionless parameter $\zeta$ defined in Suppl.~Mat. and \cite{zrake_12}.

The simulations are initialized with a Maxwellian pair plasma having uniform  density $n_0$ (including both species) and temperature $kT/mc^2=0.04$. We initialize 8 particles per cell per species and resolve the electron skin depth with 1.5 cells, so the Debye length is only marginally unresolved (see Suppl.~Mat. for convergence tests). We define $d_e=(m c^2/4\pi n_0 e^2)^{1/2}=1$ cell, and explore domain sizes from $L/d_e=250$  to $L/d_e=2000$. The speed of light is 0.45 cells/timestep. We evolve our simulations for $120\,t_{\rm L}$. Our reference box has $L/d_e=1000$. 

The energy injected by the turbulent driver eventually heats the plasma. To study the field evolution while the plasma remains in quasi-steady state, we add an artificial cooling term such that, when a particle's momentum exceeds $p_{\rm cut}$, it is reset to $p_{\rm cut}$, keeping the momentum direction unchanged.
 We choose $p_{\rm cut}/mc=0.7$ (in Suppl.~Mat. we show that our results are the same for larger $p_{\rm cut}$), so in  quasi-steady state the ratio of turbulent kinetic energy to internal energy is $\sim 0.3$ (i.e., moderately subsonic turbulence). 

\begin{figure}
\centering   \includegraphics[width=0.48\textwidth]{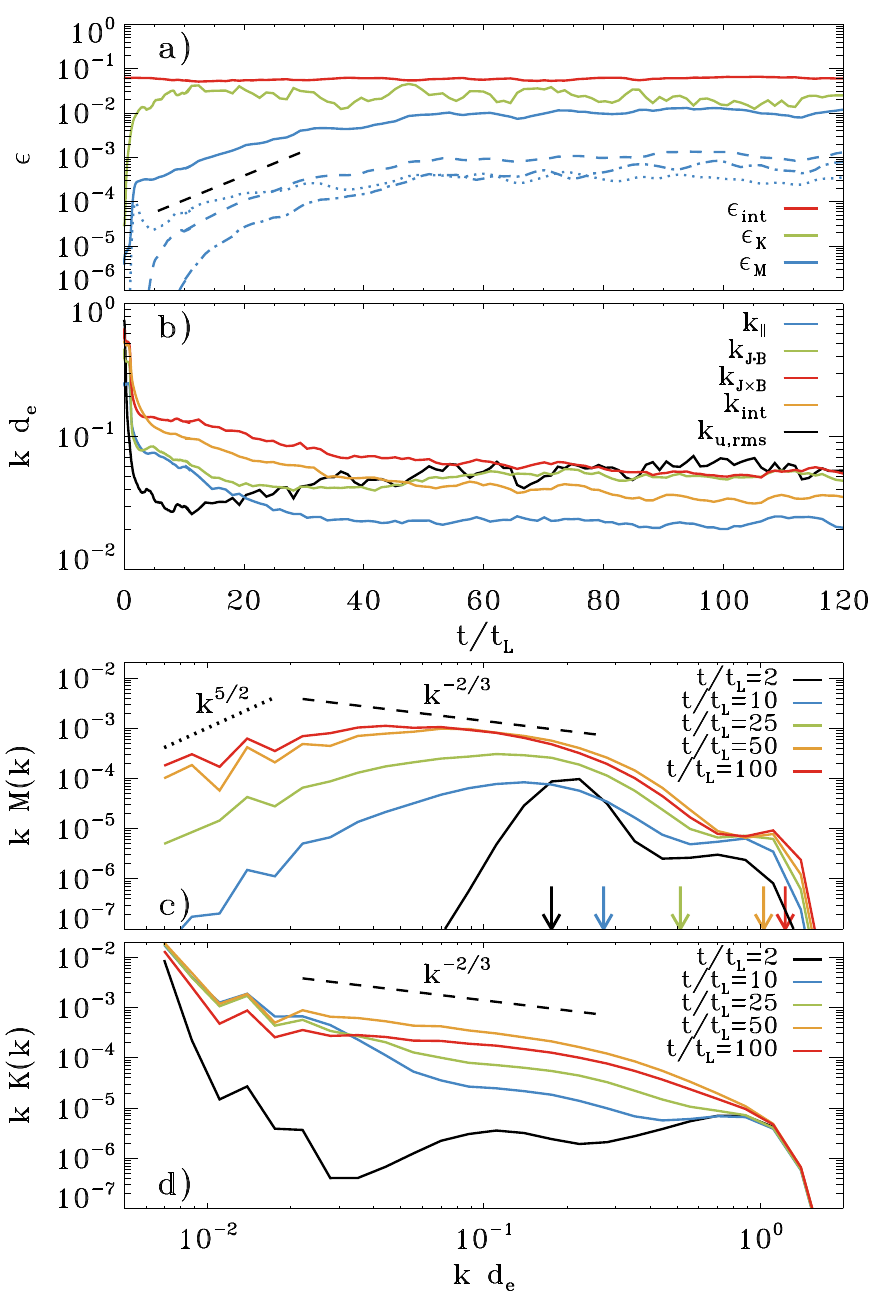}
    \caption{Time evolution of the reference simulation with $L/\de=1000$. (a) 
    Box-averaged magnetic ($\epsilon_{\rm M}$), bulk kinetic ($\epsilon_{\rm K}$), and internal ($\epsilon_{\rm int}$) energy fractions; $\epsilon_{\rm int}$ is the difference between the total energy and the bulk energy.
Dotted, dashed, and dot-dashed blue lines denote time evolution of magnetic spectral power at $k\de=0.2$, 0.04, and 0.015, respectively. The dashed black line shows $B_{\rm rms} \propto \exp(0.4 \,u_{\rm rms}t/L)$. (b) Wavevectors characterizing the magnetic field geometry (colored), and rms wavenumber of the velocity field (black), see text. (c) Angle-integrated magnetic spectrum $M(k)=\int d\Omega_k k^2 \langle {|{\bf B}}({\bf k})|^2\rangle/8 \pi$. Dotted black indicates the Kazantsev scaling $M(k)\propto k^{3/2}$ \cite{kaza_68}, whereas dashed black is the Kolmogorov scaling $M(k)\propto k^{-5/3}$ \cite{Biskamp03}. {Vertical arrows denote $\pi \langle\rho_{\rm e}^{-1}\rangle$ at various times (same color coding as the magnetic spectrum), where $\rho_{\rm e}$ is the particle Larmor radius and we average over all particles.} (d) Same as (c), but for the bulk kinetic spectrum $K(k)=\int d\Omega_k k^2 \langle {|\bf w}(\bf k)|^2\rangle$, where  $\vec w = \left[\Gamma^2_u nm/(\Gamma_u + 1)\right]^{1/2}\boldsymbol{u}$
such that $w^2 = (\Gamma_{u} - 1)nmc^2$. 
The integrals of $M(k)$ and $K(k)$ are respectively $\epsilon_{\rm M}$ and $\epsK$.}

    \label{fig:single}
\end{figure}

\begin{figure}
\centering   \includegraphics[width=0.48\textwidth]{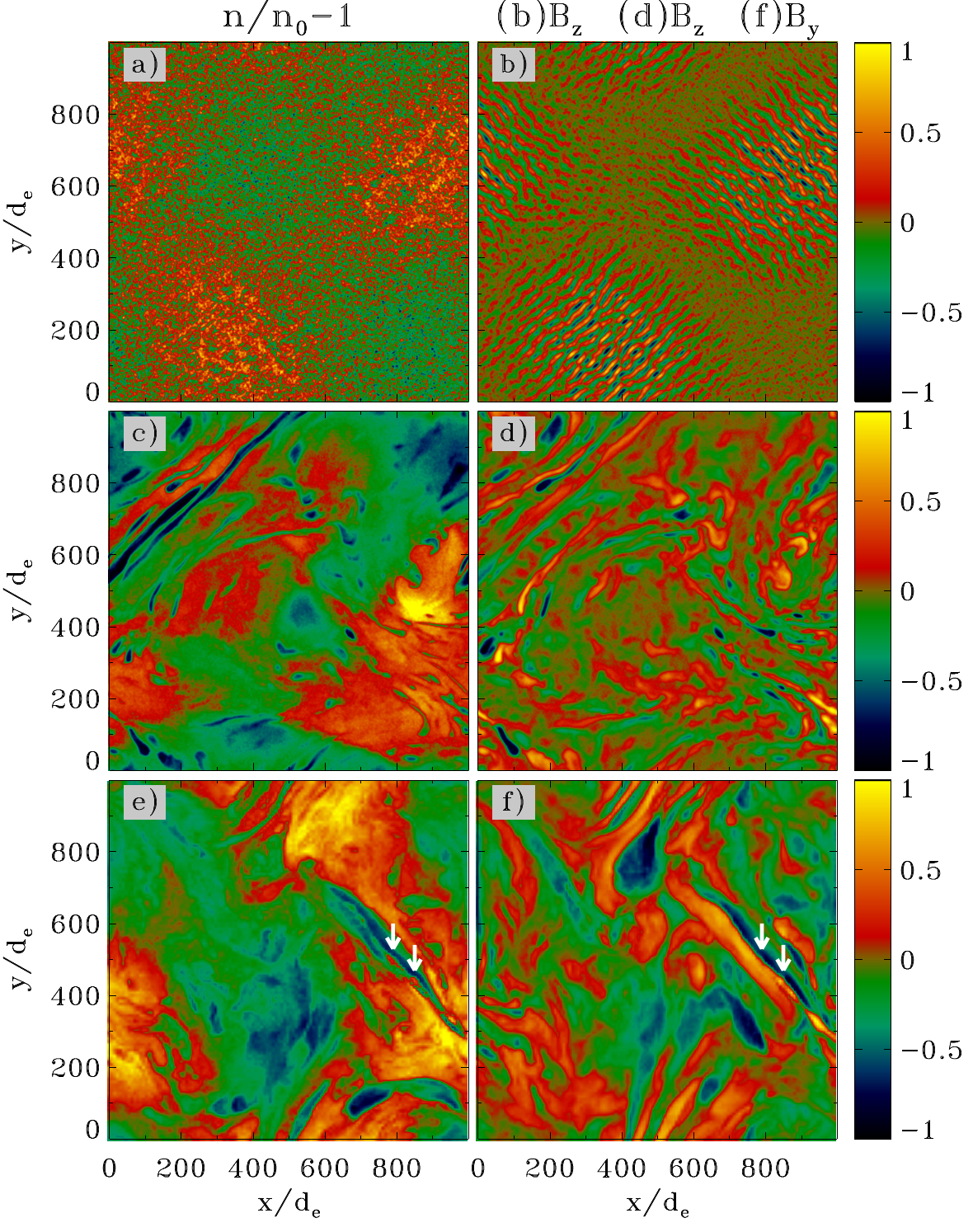}
    \caption{Representative 2D slices from the 3D reference simulation. Left: density fluctuations $n/n_0-1$. Right: $B_z$ in (b) and (d), $B_y$ in (f). Top row refers to $t/\tl=1.4$ (Weibel stage), middle row to $t/\tl=27$ (kinematic dynamo stage), bottom row to $t/\tl=120$ (saturated dynamo stage). Each panel is normalized to the maximum value of the corresponding quantity. White arrows in the bottom row point to reconnection plasmoids. See Suppl.~Mat. for a zoom-in.}
    \label{fig:fluid}
\end{figure}

{\it Results---}The magnetic field evolution, starting from an unmagnetized plasma, is illustrated in \figg{single} for our reference simulation (see a zoom-in at early times in Suppl.~Mat.), and can be divided into four stages: (i) the Weibel growth ($t/\tl\lesssim 2$), (ii) the Weibel filament merging phase ($2\lesssim t/\tl\lesssim 10$), (iii) the dynamo kinematic stage ($10\lesssim t/\tl\lesssim 40$), (iv) and the saturated state ($\ttl\gtrsim 40$). 

 Turbulence in a collisionless plasma leads to phase mixing and velocity-space anisotropies, providing free energy for the Weibel instability \cite{weibel_59}, {which, for sufficiently large domains \cite{zhou_22}, grows faster than the turbulence turnover time}. Weibel generates magnetic energy from scratch  
and saturates at a volume-averaged magnetic fraction $\epsilon_{\rm M}=\langle B^2\rangle/8\pi n_0 m c^2\simeq 3\times 10^{-4}$,  
much below the mean turbulent  energy fraction $\epsilon_{\rm K}=\langle (\Gamma_u-1)n/n_0 \rangle\simeq 0.02$, where $n$ indicates particle density and $\Gamma_u\equiv1/\sqrt{1-(u/c)^2}$, with $u$ the mean velocity (\figg{single}(a)). The Weibel growth peaks at a scale $k^{-1}\sim 5 \,\de$ (black in \figg{single}(c)), {comparable to the mean particle  Larmor radius \cite{zhou_22} at the time of Weibel saturation (vertical black arrow in \figg{single}(c)).} The fastest growing Weibel mode accounts for most of the overall field growth (compare solid and dotted blue at $t/\tl\sim 2$ in \figg{single}(a) and zoom-in in Suppl.~Mat.). The typical filamentary pattern of Weibel  fields is shown  in \figg{fluid}(b). Despite saturating at small amplitudes, the Weibel stage is essential to provide the seed fields that magnetize the plasma and feed the dynamo \cite{pusztai_20}.

The Weibel saturation is followed by a stage ($2\lesssim\ttl\lesssim 10$) where the magnetic spectrum broadens towards larger scales as a result of the merging of Weibel filaments (blue in \figg{single}(c)), while the overall magnetic energy grows only by a factor of two.
This is followed by the kinematic stage of the dynamo, from $\ttl\sim 10$ to $\ttl\sim 40$, which amplifies the magnetic energy by a factor of more than ten, up to $\epsilon_{\rm M}\simeq 10^{-2}$. 
The spectrum-integrated field grows exponentially (solid blue in  \figg{single}(a)) with rate $\gamma_B = d \ln B_{\rm rms}/dt \simeq 0.4\, u_{\rm rms}/L$ (dashed black in \figg{single}(a)). The low-$k$ end of the magnetic spectrum grows at the same rate (dashed and dot-dashed blue in \figg{single}(a), see caption; also, blue and green in \figg{single}(c)) and approaches the Kazantsev scaling, $M(k)\propto k^{3/2}$  \cite{kaza_68}. Concurrently with the magnetic growth, a Kolmogorov-like kinetic spectrum develops below the driving scale (\figg{single}(d)); its normalization falls well below the power at the driving scale, \ls{indicating  that not all fluid motions cascade to small scales, as already noticed in hybrid simulations \cite{stonge_18}.}

We further assess the field geometry by considering characteristic wavenumbers of magnetic field variation
along ($k_\parallel$) and across ($\kjdotb$, $\kjcrossb$) the field \cite{scheko_04,stonge_18,stonge_20}, see Suppl.~Mat. 
During the  kinematic dynamo stage $k_\parallel\lesssim \kjdotb \ll \kjcrossb$, see \figg{single}(b) \footnote{The same ordering is observed during the Weibel stage.}, suggestive of  fields  arranged in folded sheets \cite{stonge_20}. This is confirmed by the 2D slice in \figg{fluid}(d).

The kinematic dynamo phase terminates at $t/\tl\sim 40$, and the system settles into a quasi-steady saturated stage. Here, the magnetic energy is about
half of the turbulent kinetic energy (compare solid blue and green in \figg{single}(a)), regardless of the plasma temperature (see Suppl.~Mat.). The magnetic spectrum peaks near $k\simeq 0.04\, d_e^{-1}\simeq 12 \pi/L$ (red in \figg{single}(c)) and it is shallower than  the Kazantsev scaling at lower $k$. The spectral peak can be approximated by the integral-scale wavenumber,
\begin{equation}
k_{\rm int}=\frac{\int M(k) dk}{\int k^{-1} M(k) dk} \, ,
\end{equation}
whose time evolution is shown by the yellow line in \figg{single}(b). The kinetic  spectrum displays a Kolmogorov-like scaling at $k\gtrsim 6 \pi/L$, with a normalization that is a factor of ten lower than the power at the driving scale (red in \figg{single}(d)).
In the saturated stage, $k_\parallel\ll \kjdotb \sim \kjcrossb$ (\figg{single}(b)). This is suggestive of folded magnetic fields organized into flux ropes, a natural outcome of the tearing disruption of current sheets \cite{galishnikova_22}. In fact, \figg{fluid}(e) shows the presence of over-dense plasmoids (indicated by the white arrows) within a current sheet where $B_y$ switches polarity (\figg{fluid}(f)). In the saturated stage the rms wavenumber of the kinetic spectrum,
\begin{equation}
k_{\rm u,rms}=\left(\frac{\int k^2 K(k) dk}{\int K(k) dk}\right)^{1/2},
\end{equation}
approaches $\sim \kjdotb \sim \kjcrossb$, 
indicating flows on the flux-rope scale, as observed in high-resolution MHD simulations and argued to be evidence of tearing modes  \cite{galishnikova_22}.

\begin{figure}
\centering   \includegraphics[width=0.48\textwidth]{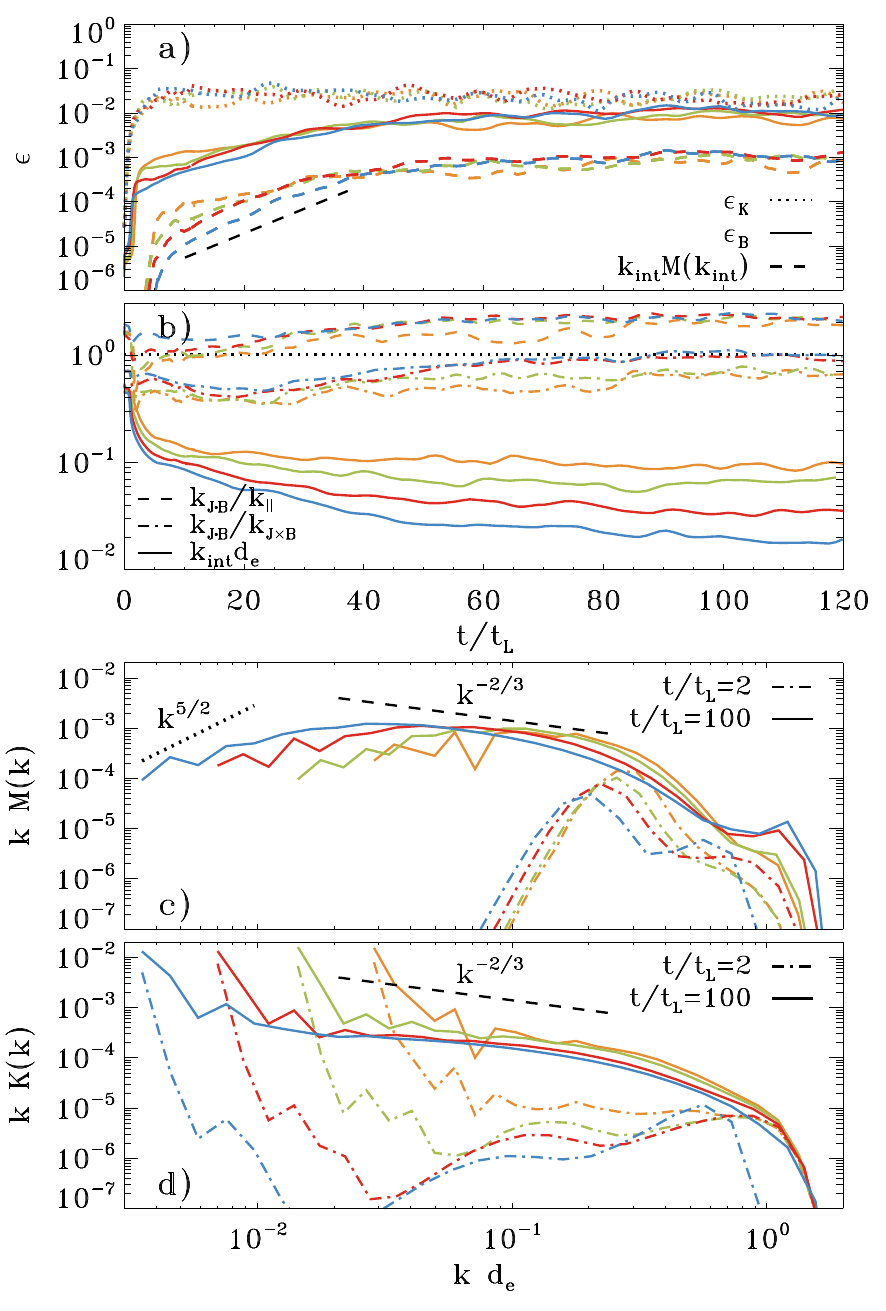}
    \caption{Dependence on box size: $L/\de=250$ (yellow), $L/\de=500$ (green), $L/\de=1000$ (red), and $L/\de=2000$ (blue). (a) Kinetic ($\epsK$; dotted lines) and magnetic ($\epsM$; solid lines) energy fractions. Dashed colored lines track the growth of magnetic spectral power at large scales: $k\de=0.09$, 0.06, 0.04, and 0.03 from smallest to largest boxes. The dashed black line shows $B_{\rm rms} \propto \exp(0.4 \,u_{\rm rms}t/L)$. (b) Time evolution of  $\kintde$ (solid), $k_{\rm J \cdot B}/k_\parallel$ (dashed) and $k_{\rm J \cdot B}/k_{\rm J \times B}$ (dot-dashed). Magnetic (c) and bulk kinetic (d) power spectra right after the end of Weibel growth (dot-dashed) and in the quasi-steady state at late times (solid). Dotted black in (c) is the Kazantsev $M(k)\propto k^{3/2}$ \cite{kaza_68}, whereas dashed black in (c) and (d) is the Kolmogorov $M(k)\propto k^{-5/3}$ \cite{Biskamp03}.}
    \label{fig:box}
\end{figure}

\figg{box} shows the dependence of our results on the box size.
In larger boxes the Weibel growth peaks at lower $kd_e$ (dot-dashed lines in \figg{box}(c)), and the field saturates at smaller amplitudes (see \figg{box}(a) and zoom-in in Suppl.~Mat.), in agreement with \cite{zhou_22}.

Beyond the Weibel phase, our results show excellent convergence for sufficiently large boxes. In the kinematic stage, the low-$k$ end of the magnetic spectrum grows exponentially with rate $\gamma_B \simeq 0.4 \, u_{\rm rms}/L$ for $L/\de \geq 500$ (solid and dashed  in \figg{box}(a)). The presence of flux ropes --- as inferred from $\kjdotb\sim\kjcrossb$ at late times --- is manifested for large enough boxes, $L/\de\geq 1000$ (red and blue dot-dashed  in \figg{box}(b)). In the saturated stage, the magnetic energy  is roughly half of the turbulent kinetic energy, regardless of box size. The magnetic spectrum in the saturated stage is broad, with $kM(k)$ extending nearly over two decades (for $L/d_e=2000$) and peaking at $\kint\simeq 12 \pi/L$ (solid in \figg{box}(c)). For our largest domain, this corresponds to a wavelength $\simeq 0.16 \, L\simeq 300\,d_e$, i.e., much larger than kinetic plasma scales. 

The integral scale of the magnetic spectrum at saturation is consistent with a dynamo hindered by fast magnetic reconnection. We define the reconnection timescale $t_{\rm rec} = (\beta_{\rm rec} \delta v_A \kint)^{-1}$, where $\beta_{\rm rec} \sim 0.1$ \cite{Comisso16,Cassak17} is the fast collisionless reconnection rate and $\delta v_A$ is the Alfv{\'e}n speed associated with $B_{\rm rms}$. Comparing this with the turnover time $t_{L} = (L/2\pi)/u_{\rm rms}$, one infers $\kint \sim (2\pi/\beta_{\rm rec} L) (u_{\rm rms}/\delta v_A)$. Since $\delta v_A \sim u_{\rm rms}$ at saturation, fast magnetic reconnection gives $\kint \sim 20 \pi/L$, which is close to the value observed in the simulations.

\begin{figure}
\centering   \includegraphics[width=0.48\textwidth]{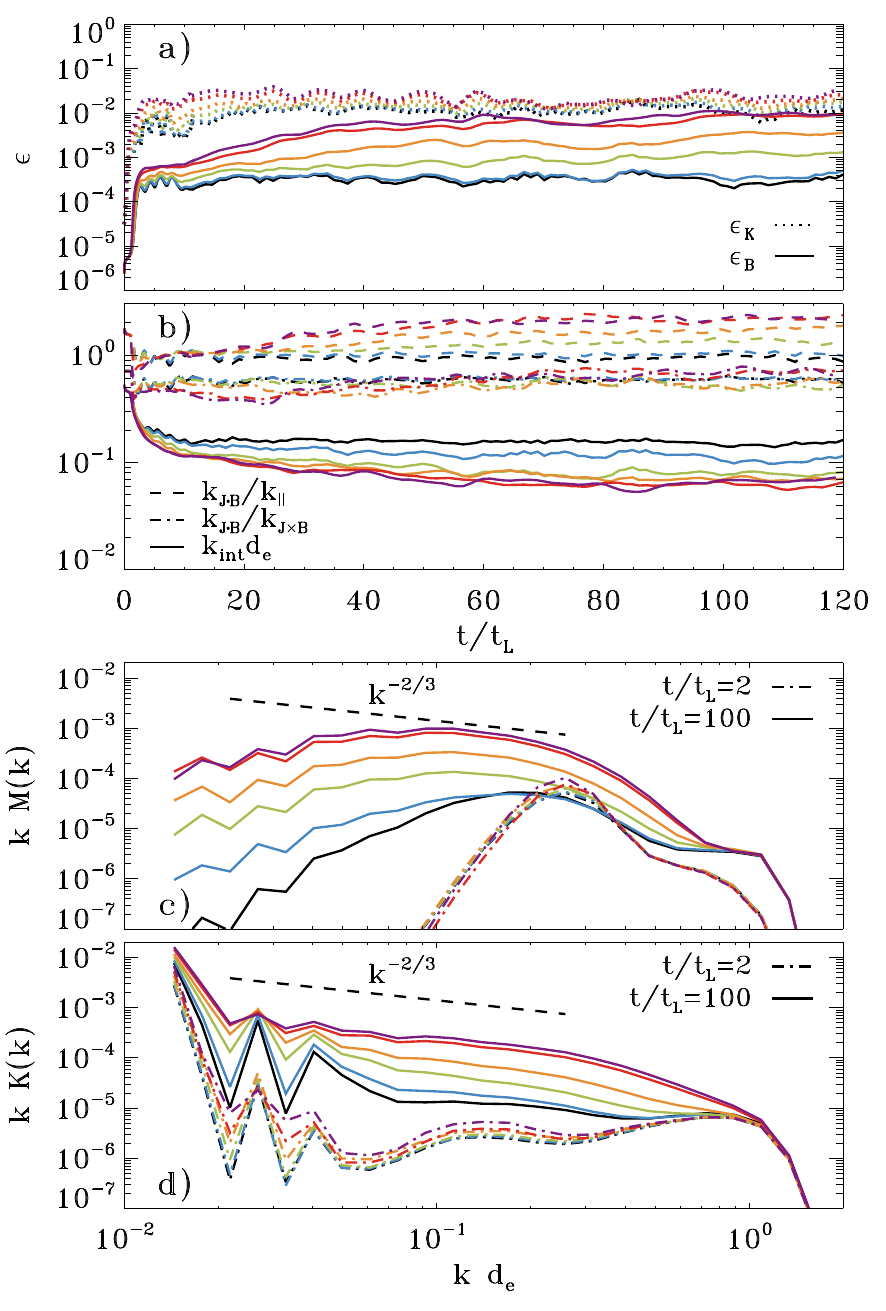}
    \caption{Same as \figg{box}, but exploring the dependence on the driver, for a box with $L/\de=500$. We vary  $\zeta$ (see Suppl.~Mat. and \cite{zrake_12}), with $\zeta=0$ for  compressive modes and $\zeta=1$ for  vortical modes. We show $\zeta=0$ (black), 0.1 (blue), 0.2 (green), 0.3 (yellow), 0.5 (red) and 1 (purple).}
    \label{fig:compress}
\end{figure}

Lastly, we demonstrate that efficient dynamo growth requires that a significant fraction of the turbulent energy is in vortical motions. In \figg{compress} we consider a box with $L/\de=500$ and vary the parameter $\zeta$ defined in Suppl.~Mat. following \cite{zrake_12}, with $\zeta=0$ for purely compressive modes and $\zeta=1$ for purely vortical modes (as employed so far).
We find that the Weibel phase proceeds similarly regardless of $\zeta$, and the magnetic and kinetic spectra at $t/\tl=2$---dot-dashed  in \figg{compress}(c) and (d), respectively---are nearly independent of $\zeta$. Yet, the subsequent evolution is markedly different. For $\zeta=0$, the magnetic energy fraction stays nearly constant after the Weibel saturation (\figg{compress}(a), black solid). The magnetic spectrum broadens towards lower $k$ as a result of filament merging (\figg{compress}(c), compare solid and dot-dashed black), but the integral-scale wavenumber,  $\kint$, drops by only a factor of two from $t/\tl=2$ (the time of Weibel saturation) to the end of the simulation (\figg{compress}(b), solid black). 
The case $\zeta=0.1$ resembles the purely compressive case, $\zeta=0$, whereas evidence of dynamo growth appears for $\zeta=0.2$ or larger---the magnetic energy fraction increases after the Weibel phase and  $\kint$ significantly drops. For $\zeta\geq0.2$, the magnetic spectra at saturation have nearly the same shape, albeit with different normalization (\figg{compress}(c)). We conclude that $\zeta\geq0.2$ is necessary for some degree of dynamo action. Yet, only for $\zeta=0.5$ (red lines) or above the magnetic energy reaches near-equipartition with the turbulent kinetic energy.  \ls{Earlier MHD simulations \cite{federrath_11,federrath_21} reached the same conclusion, i.e., compressive modes are unable to drive dynamo growth. The agreement of our results with MHD simulations confirms that the field evolution seen in our vortical runs after Weibel saturation can be convincingly attributed to MHD-like dynamo action (as opposed to a nonlinear evolution of Weibel-generated fields, which also occurs in the purely compressive case).}

{\it Conclusions}---We drive turbulence in a collisionless, initially unmagnetized pair plasma,  mimicking the effect of turbulent  gravitational forces generated by the build-up of the large-scale structure. As shown in \cite{weibel_59,Davidson72,Medvedev99,Silva03,Medvedev06,Silva21,califano_98,califano_06,bret_08,pucci_21,zhou_22}, the Weibel instability generates seed magnetic fields, which however saturate well below equipartition with the turbulent kinetic energy. If most of the turbulent energy is in vortical motions, Weibel saturation is followed by dynamo amplification and the turbulent dynamo brings the field energy up to near-equipartition with the turbulent kinetic energy. For large boxes, the integral scale of the magnetic spectrum is roughly an order of magnitude smaller than the driving scale, consistent with dynamo saturation controlled by fast collisionless reconnection. 


Our results have important implications for weakly collisional plasmas in our Universe. 
Observations of Faraday rotation and synchrotron emission \cite{bonafede_10} show that ICM fields have strengths $\sim \mu{\rm G}$, with energy density comparable to that of turbulent motions \cite{hitomi_16}. The magnetic spectrum is inferred to peak on scales $\sim 10\,\rm kpc$ \cite{vacca_12,govoni_17}, a factor of 10 smaller than the turbulence driving scale $L\sim 100\,\rm kpc$, in good agreement with our findings. 
We caution, though, that the scale separation, $L/d_e$, in our simulations is \ls{many} orders of magnitude smaller than in the ICM. {Further work is needed to confirm that our findings hold for even larger values of $L/d_e$ and to clarify the interplay between the electron Weibel instability, the ion Weibel instability, and the Biermann battery \cite{biermann_50,schoeffler_14,schoeffler_16} in the realistic case of electron-ion plasmas.}

\begin{acknowledgments}
L.S. acknowledges support by the Cottrell Scholar Award. L.S. and R.G. are 
 supported by NASA ATP Grant No.~80NSSC20K0565. L.C. acknowledges support by DOE DE-SC0021254.
 This research was facilitated by Multimessenger Plasma Physics Center (MPPC), NSF grant PHY-2206609. 
Simulations were performed on NASA Pleiades. 
\end{acknowledgments}

\bibliographystyle{apsrev}
\bibliography{magneto.bib}

\section{Supplemental Material}

\subsection{Additional Details on Setup and  Diagnostics}
\ls{We perform 3D PIC simulations with TRISTAN-MP \citep{buneman_93, spitkovsky_05} in a periodic cube of size $L^3$. We employ a first-order particle shape, the standard Yee mesh \cite{yee_66} for solving Maxwell's equations,  and the Boris algorithm \cite{Boris70} for pushing charged particles via the Lorentz force. The electromagnetic fields are extrapolated to the particle positions via a tri-linear interpolation function (linear in each dimension). Electric currents are deposited on the numerical grid using the zigzag scheme of \cite{umeda_03} and smoothed with 10 passes of a 3-point (1-2-1) digital filter acting on each axis.}

We drive six modes on the scale of the box, with ${\bf k} L/2\pi =(\pm 1,0,0),\, (0,\pm 1,0),\, (0,0,\pm 1)$. We drive turbulence by adding a charge-independent external acceleration term ${\bf a}({\bf x},t)$ to the equations of motion of PIC macro-particles, which acts in addition to the Lorentz force. The Fourier modes $\tilde{\bf a}({\bf k},t)$  of the driving field are advanced according to an Ornstein--Uhlenbeck process, which consists of a restoring force together with a complex-valued random-walking term $d\tilde{\bf W}({\bf k},t)$:
$$
d\tilde{\bf a}({\bf k},t)=-\gamma_{\rm corr}\,dt\, \tilde{\bf a}({\bf k},t)+\sqrt{\gamma_{\rm corr} P_k} {\bf T}({\bf k})\cdot d\tilde{\bf W}({\bf k},t)
$$
\ls{Here, the decorrelation rate is chosen to be $\gamma_{\rm corr}\simeq 1.5/\tl$, such that the de-correlation time is comparable to the turnover time $\tl=(L/2\pi)/u_{\rm rms}$.} The random-walking term $d\tilde{\bf W}({\bf k},t)$ is a complex random number whose real and imaginary parts are uniformly distributed $\in [-1/2,1/2]$. The projection operator 
$$
{\bf T}({\bf k})=(1-\zeta){\bf T}^\parallel({\bf k})+\zeta {\bf T}^\perp({\bf k})
$$
is applied to the vector deviate $d\tilde{\bf W}({\bf k},t)$ in order to select compressive or vortical driving, according to the parameter $\zeta$, with $\zeta=0$ for purely compressive modes and $\zeta=1$ for purely vortical modes \cite{zrake_12}. More explicitly, 
\begin{eqnarray}
{\bf T}^\parallel({\bf k}) \cdot d\tilde{\bf W}({\bf k},t) & = & [\hat{\bf k} \cdot d\tilde{\bf W}({\bf k},t)] \,\hat{\bf k} \, ,\\
{\bf T}^\perp({\bf k}) \cdot d\tilde{\bf W}({\bf k},t) & = & d\tilde{\bf W}({\bf k},t) -[\hat{\bf k} \cdot d\tilde{\bf W}({\bf k},t)] \,\hat{\bf k} \, .
\end{eqnarray}
The normalization $P_k$ is the same for all modes and accounts for the different degrees of freedom of compressive vs. solenoidal driving, so we choose $P_{k}\propto[(1-\zeta)^2+2\zeta^2]^{-1}$. The value of $P_k$ is such that the  target rms velocity is $u_{\rm rms}\sim 0.2 \,c$. \ls{This is sufficiently small, as compared to the speed of light, that our simulations are reasonably in the regime of non-relativistic turbulent dynamo.}

The spatial realization of the external acceleration is obtained by taking the real part of the Fourier mode superposition,
\begin{equation}
{\bf a}({\bf x},t)=\Re\left[\Sigma_{\bf k} \tilde{\bf a}({\bf k},t) \exp(i{\bf k} \cdot {\bf x})\right] \, ,
\end{equation}
where the sum is over the six modes driven in the box.

Finally, we formally define the characteristic wavenumbers of magnetic-field variation
along ($k_\parallel$) and across ($\kjdotb$, $\kjcrossb$) itself, as in \cite{scheko_04,stonge_18,stonge_20}. We define
\begin{eqnarray}
k_\parallel&=&\left(\frac{\langle|{\bf B\cdot {\bf \nabla}} {\textbf B}|^2\rangle}{\langle B^4 \rangle} \right)^{1/2} \, , \\
\kjdotb&=&\left(\frac{\langle|{\bf B\cdot \tilde{J}}|^2\rangle}{\langle B^4 \rangle} \right)^{1/2} \, , \\
\kjcrossb&=&\left(\frac{\langle|{\bf B\times \tilde{J}}|^2\rangle}{\langle B^4 \rangle} \right)^{1/2} \, ,
\end{eqnarray}
where ${\bf \tilde J}={\bf \nabla }\times {\bf B}$. $\kjcrossb$  captures the field reversals in folds; $\kjdotb$ measures the variation of the field in the direction orthogonal to both $\bf B$ and $\tilde{\bf J} \times \bf B$.

\begin{figure}[!t]
\centering   \includegraphics[width=0.48\textwidth]{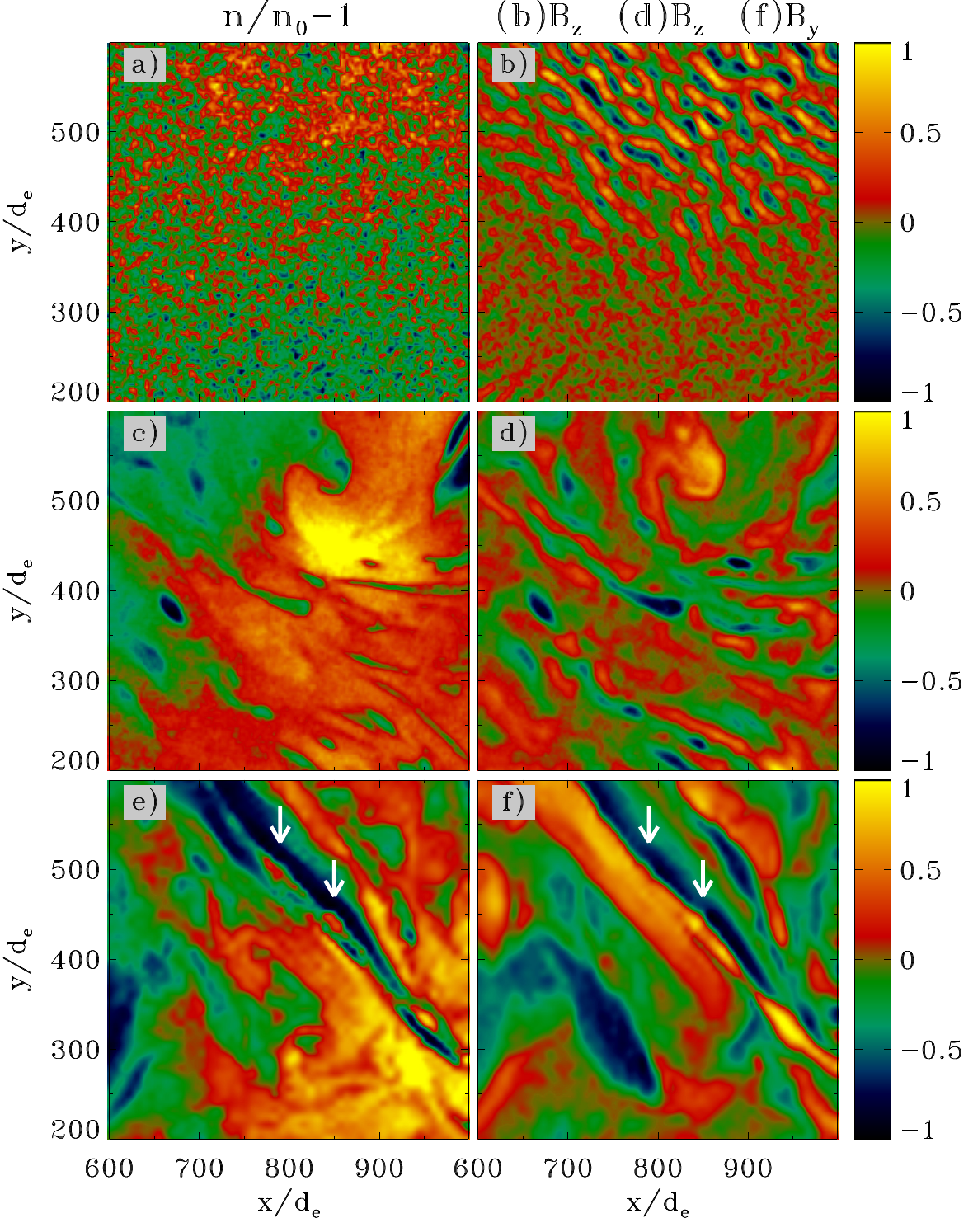}
    \caption{Same as Fig.~2 in the main paper, but zooming in on a smaller spatial range. White arrows in the bottom row point to plasmoid-like structures within a current sheet where $B_y$ changes direction.}
    \label{fig:fluidzoom}
\end{figure}

\begin{figure}[!t]
\centering   \includegraphics[width=0.48\textwidth]{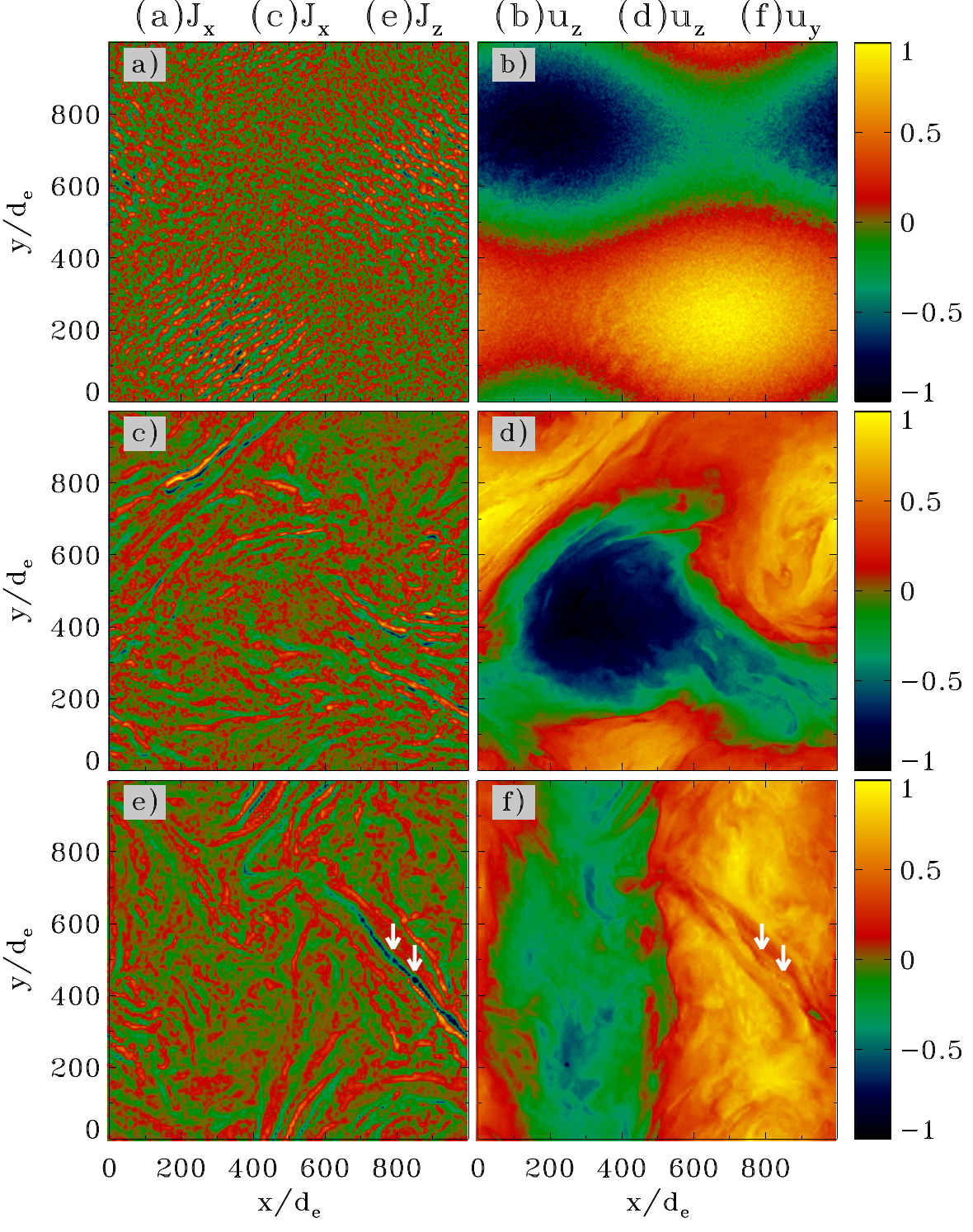}
    \caption{Same as Fig.~2 in the main paper, but for different fluid quantities. We show the electric current density in the left column ($J_x$ in (a) and (c), $J_z$ in (e)) and the fluid velocity in the right column ($u_z$ in (b) and (d), $u_y$ in (f)).}
    \label{fig:fluidex}
\end{figure}
\subsection{Fluid Structure}

In \figg{fluidzoom} and \figg{fluidex}, we present the same time snapshots as employed for Fig.~2 in the main paper. \figg{fluidzoom} is a zoom-in view of the same quantities presented in Fig.~2 of the main paper, which helps identify the characteristic signatures of Weibel filaments in the Weibel exponential growth phase (\figg{fluidzoom}(b)) and of  reconnection plasmoids in the saturated stage (\figg{fluidzoom}(e),  indicated by the white arrows). \figg{fluidex} displays one component of the electric current (left) and one component of the fluid velocity (right). \figg{fluidex}(e) demonstrates that the  plasmoid-like structures identified in Fig.~2(e) of the main paper and \figg{fluidzoom}(e) here are indeed localized within a sheet of intense $J_z$, further corroborating their identification as reconnection plasmoids. The right column of \figg{fluidex} shows that the velocity field is generally much smoother than the magnetic field (compare with the right column of Fig.~2 in the main paper), as already noted in hybrid simulations \cite{stonge_18} and consistent with the fact that the magnetic spectrum peaks at smaller scales than the velocity  spectrum. At the current sheet in \figg{fluidex}(e), the magnetic shear is much stronger than the velocity shear (the $u_y$ velocity in \figg{fluidex}(f) is nearly continuous across the magnetic discontinuity).

\begin{figure}
\centering   \includegraphics[width=0.48\textwidth]{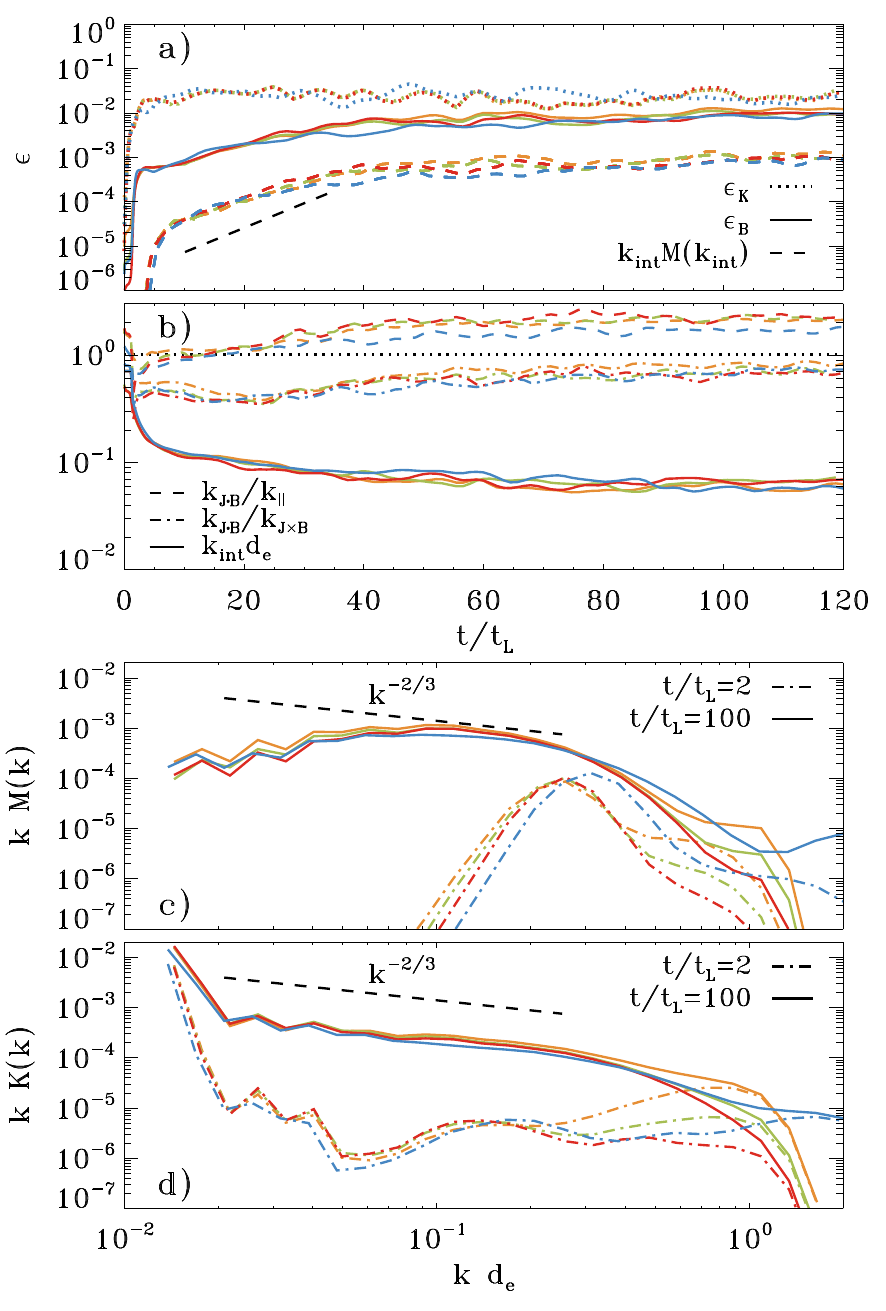}
    \caption{
    Same as Fig.~3 in the main paper, but exploring numerical convergence, for a box size $L/\de=500$. Yellow, green, and red lines employ 1.5 cells per electron skin depth (our standard resolution) and vary in the number of particles per cell per species: 2 (yellow), 8 (green; our reference case), and 32 (red). The blue curve employs 8 particles per cell per species and a higher resolution of 3 cells per electron skin depth.    
    Dashed lines in (a) track the growth of magnetic spectral power at $k\de=0.06$.
    }
    \label{fig:ppc}
\end{figure}

\begin{figure}
\centering   \includegraphics[width=0.48\textwidth]{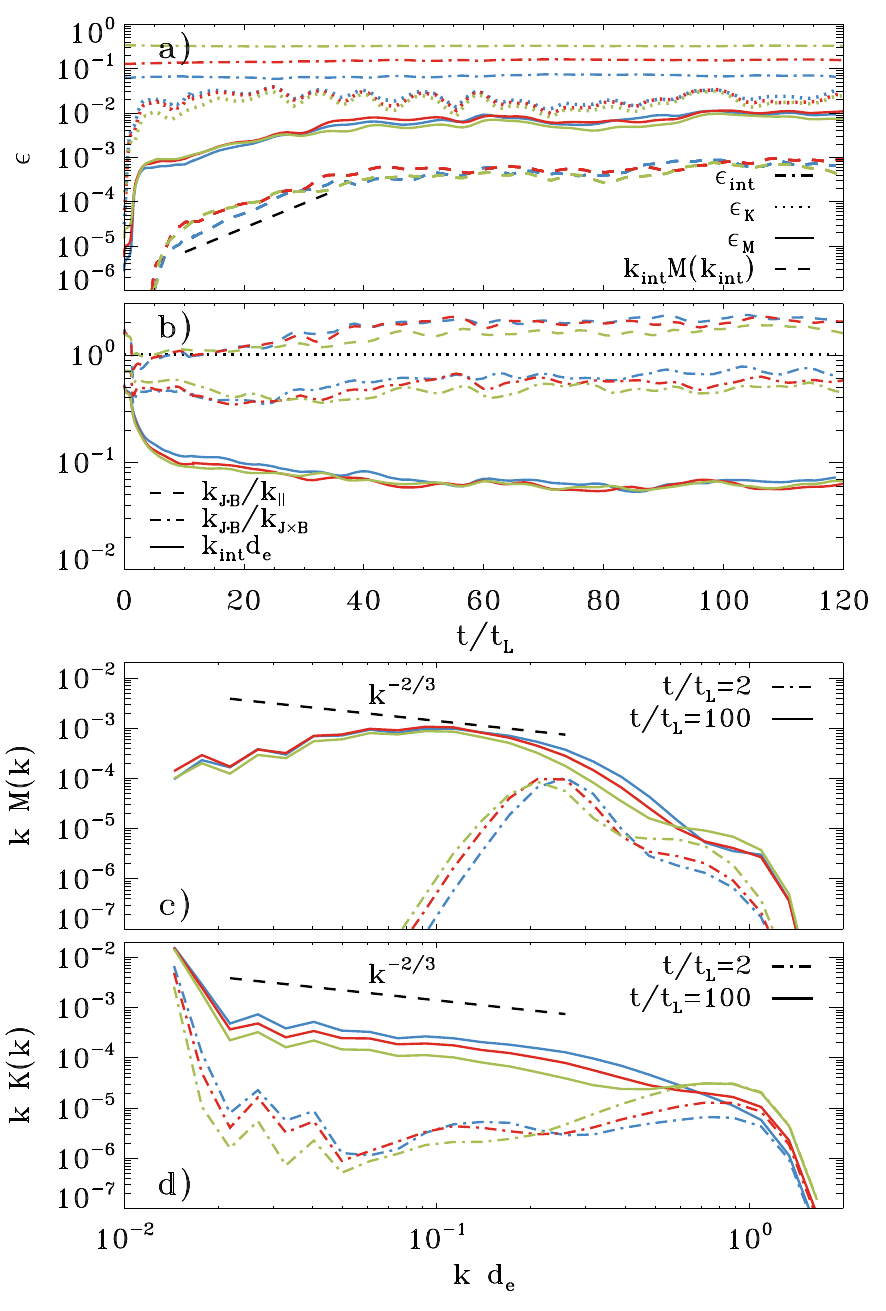}
    \caption{Same as Fig.~3 in the main paper, but exploring the dependence on the initial temperature and the cutoff momentum, $p_{\rm cut}$, for a box size $L/\de=500$.  Blue is initialized with $kT/mc^2=0.04$ and uses $p_{\rm cut}=0.7mc$ (our reference choices), red employs  $kT/mc^2=0.08$ and $p_{\rm cut}=1.0mc$, and green uses $kT/mc^2=0.2$ and $p_{\rm cut}=1.5mc$.
    In (a), dashed lines track the growth of magnetic spectral power at $k\de=0.04$, while dot-dashed lines illustrate the time evolution of the internal energy fraction.
}
    \label{fig:ptarget}
\end{figure}

\begin{figure}
\centering   \includegraphics[width=0.48\textwidth]{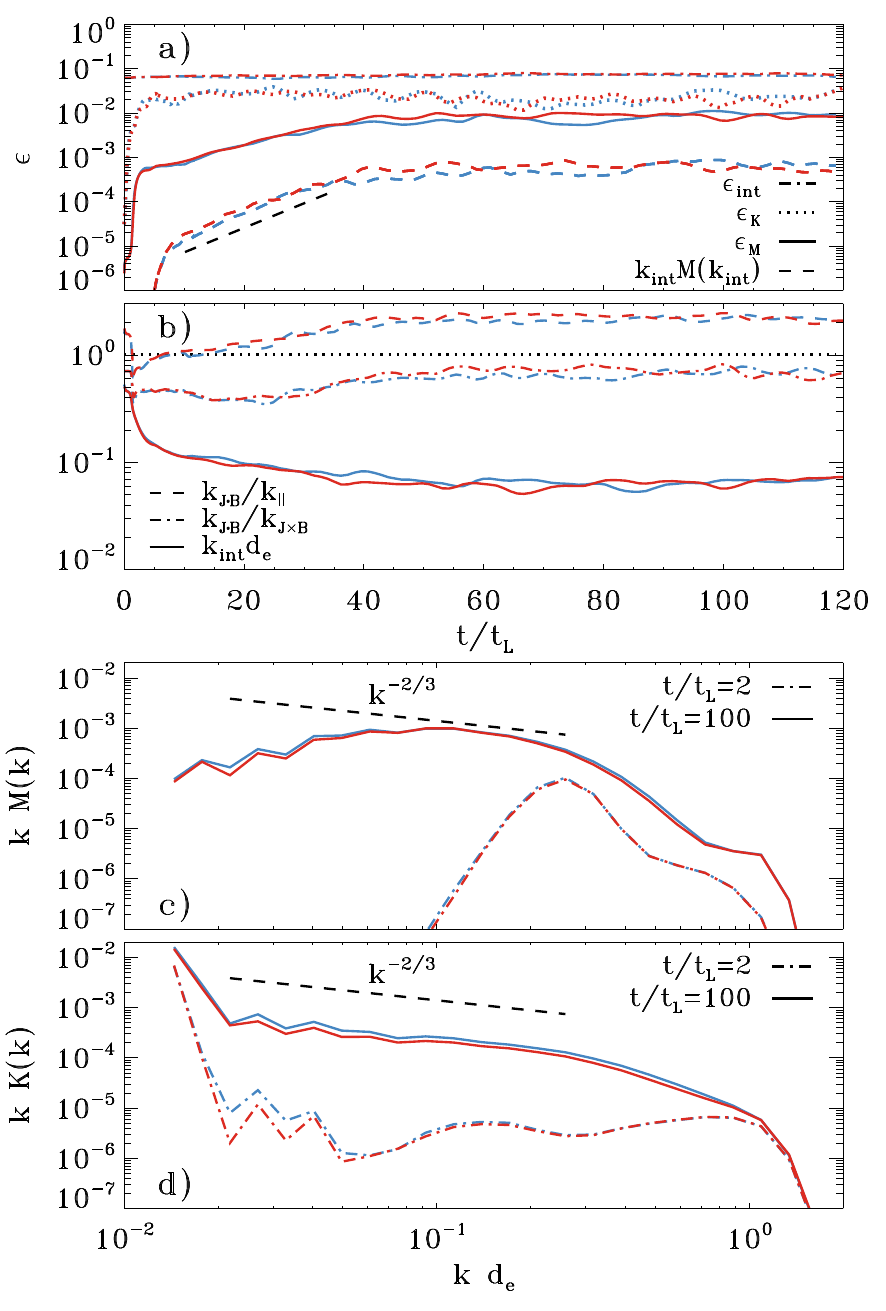}
    \caption{\ls{Same as Fig.~3 in the main paper, but exploring the dependence on the cooling strategy, for a box size $L/\de=500$.  Blue is evolved in the standard way (particles exceeding $p_{\rm cut}=0.7mc$ are reset exactly at $p_{\rm cut}$), while red is our new test, where $p_{\rm cut}=0.8mc$ but if a particle exceeds this value, its momentum is randomly reset in the range between $2p_{\rm cut}/3$ and $p_{\rm cut}$.
    In (a), dashed lines track the growth of magnetic spectral power at $k\de=0.04$, while dot-dashed lines illustrate the time evolution of the internal energy fraction.}
}
    \label{fig:ptarget2}
\end{figure}

\subsection{Numerical Convergence}
\label{sec:converge}
\figg{ppc} shows the same quantities as Fig.~3 in the main paper and demonstrates that our results are numerically converged. We find no significant dependence on the number of particles per cell (2 per species in yellow, 8 in green, 32 in red). 

Regarding spatial resolution, the blue line employs 8 particles per species per cell and a spatial resolution  of $d_e=2$ cells---corresponding to $\simeq 3$ cells per electron skin depth---that is twice as high as  our fiducial choice. The Weibel phase proceeds similarly for $d_e=1$ and $d_e=2$, with similar saturation levels for the magnetic energy. 
The only notable difference is the fact that the magnetic spectrum near the end of the Weibel phase (dot-dashed lines in \figg{ppc}) peaks at $k \de\simeq 0.25$ for our fiducial runs, whereas it peaks at $k \de\simeq 0.3$ for the higher resolution case (so, a moderate $\simeq 20\%$ increase when increasing the resolution by a factor of two). This small difference carries over to the high-$k$ end (near $kd_e\simeq 0.6$) of the magnetic spectrum in the saturated stage, which extends to slightly larger $k$  for the higher resolution case. 

In the saturated stage the overall magnetic content (solid in \figg{ppc}(a)), the integral scale of the magnetic spectrum (solid in \figg{ppc}(b)), and the whole low-$k$ end of the magnetic spectrum (solid in \figg{ppc}(c)) are clearly independent of the number of particles per cell and of the spatial resolution. We therefore regard our conclusions as numerically robust.

\subsection{Dependence on the Plasma Temperature and the Cutoff Momentum}
In \figg{ptarget} we change the initial plasma temperature and the cutoff momentum $p_{\rm cut}$ used to establish the quasi-steady state (see main paper for details). This leads to different ratios between the plasma internal energy and the turbulent kinetic energy -- i.e., we explore the dependence of our results on the turbulence Mach number. In the hottest case with $kT/mc^2=0.2$, the steady-state ratio of turbulent kinetic energy to internal energy is $\sim 0.06$, corresponding to a turbulent Mach number $\sim 0.2$. \figg{ptarget} shows that our conclusions do not depend on the plasma thermal properties. In particular, the steady-state magnetic energy is always roughly half of the turbulent kinetic energy (solid in \figg{ptarget}(a)), regardless of the plasma temperature. The integral scale of the magnetic power spectrum is also unchanged (solid in \figg{ptarget}(b)). The only  difference is that hotter plasmas lead to a slightly lower high-$k$ cutoff of the magnetic spectrum (solid in \figg{ptarget}(c)), likely because in our definition of $d_e$ we have omitted corrections for relativistic temperatures.
Hotter plasmas also display a larger drop in the normalization of the Kolmogorov-like range ($0.03\lesssim kd_e\lesssim 0.3$) of the saturated kinetic spectrum (solid in \figg{ptarget}(d)), as compared to the power at the driving scale. 

\ls{To confirm the robustness of our findings with respect to the cooling strategy, we have performed one additional test. In our usual strategy, once a particle exceeds $p_{\rm cut}=0.7 mc$, its momentum is reset exactly at the threshold value. In our new test, we select a slightly larger $p_{\rm cut}=0.8mc$ and we prescribe that, once a particle exceeds this value, its momentum is randomly reset in the range between $2p_{\rm cut}/3$ and $p_{\rm cut}$. The results of this new test are shown in \figg{ptarget2}, where we demonstrate that they are in excellent agreement with our standard strategy.}

\subsection{Zoom-in on the Weibel Phase and the Filament Merging Stage}
\begin{figure}
\centering   \includegraphics[width=0.47\textwidth]{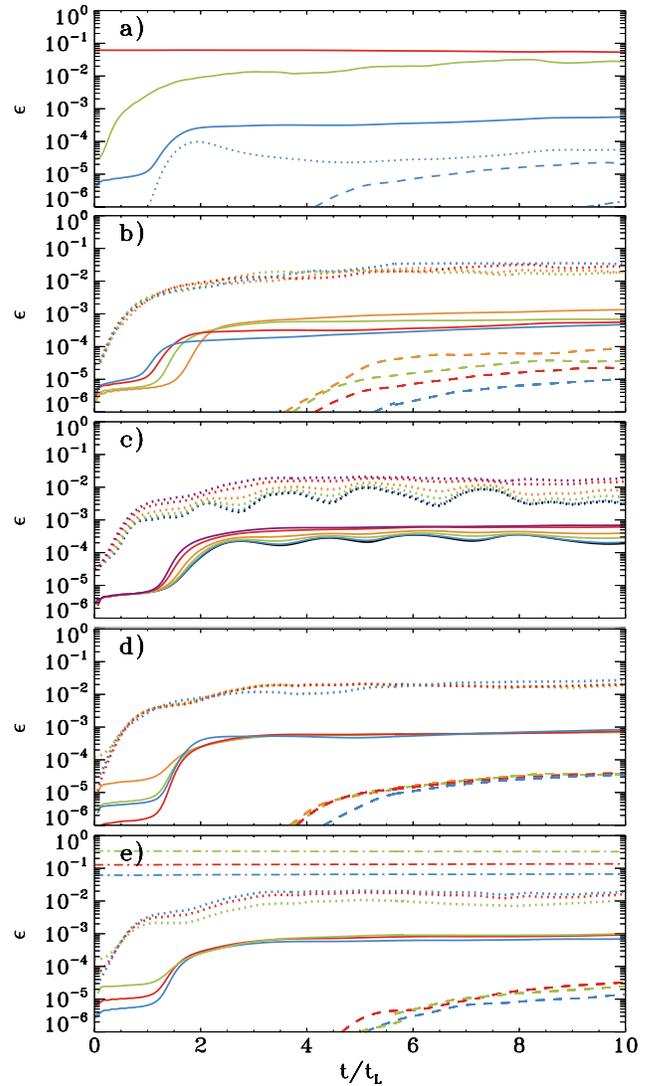}
    \caption{Zoom-in on the evolution at early times to emphasize the Weibel phase and the filament merging stage. We display a zoom-in of panels (a) of Fig.~1, 3 and 4 in the main paper, and of \figg{ppc} and \figg{ptarget} in the Suppl.~Mat. We show: (a) Magnetic (blue), bulk kinetic   (green), and internal  (red) energy fractions. Dotted, dashed, and dot-dashed blue lines denote the time evolution of magnetic spectral power at $k\de=0.2$, 0.04, and 0.015, respectively. (b) Magnetic (solid) and bulk kinetic (dotted) energy fractions for $L/\de=250$ (yellow), $L/\de=500$ (green), $L/\de=1000$ (red), and $L/\de=2000$ (blue).  (c) Magnetic (solid) and bulk kinetic (dotted) energy fractions, for $\zeta=0$ (black), 0.1 (blue), 0.2 (green), 0.3 (yellow), 0.5 (red) and 1 (purple). (d) Magnetic (solid) and bulk kinetic (dotted) energy fractions for our convergence studies. Yellow, green, and red  employ $d_e=1$ and different particles per cell per species: 2 (yellow), 8 (green), and 32 (red). The blue curve employs 8 particles per cell per species and $d_e=3$.  (e) Magnetic (solid), bulk kinetic (dotted), and internal (dot-dashed) energy fractions for different plasma temperatures. Blue has $kT/mc^2=0.04$ and $p_{\rm cut}=0.7$ (our reference choices), red has  $kT/mc^2=0.08$ and  $p_{\rm cut}=1.0$, and green has $kT/mc^2=0.2$ and  $p_{\rm cut}=1.5$. In (b), (d) and (e), dashed lines track the growth of magnetic spectral power at large scales.
}
    \label{fig:weibel}
\end{figure}

\figg{weibel} provides a zoom-in on the evolution at early times to emphasize the Weibel phase and the filament merging stage. For a simple shear flow, these phases have been extensively characterized by \cite{zhou_22}. We summarize here the salient points conveyed by \figg{weibel}.

It is apparent that the bulk kinetic energy reaches a quasi-steady state before the magnetic energy grows exponentially (at $t/\tl\sim 2$) as a result of the Weibel instability. Thus, at the time of Weibel growth, the plasma bulk motions are already well developed. \figg{weibel}(a) shows, as anticipated in the main paper, that  during the Weibel stage the fastest growing mode (dotted blue) accounts
for most of the overall field growth (solid blue). 

In \figg{weibel}(b), where we investigate the dependence on $L/d_e$, it is apparent that for larger boxes the Weibel exponential growth occurs earlier (in units of $\tl$) and the magnetic energy saturates at lower levels. Fig.~3(c) in the main paper shows that the magnetic spectral peak shifts to smaller $kd_e$ for larger systems. These three trends are in full agreement with the work of \cite{zhou_22}. In order to account for the $L/d_e$-dependence of the Weibel onset time, the spectra shown with dot-dashed lines in Fig.~3(c) and (d) of the main paper are measured at $t/\tl\simeq 2.5$ for $L/\de=250$, $t/\tl\simeq 2$ for $L/\de=500$, $t/\tl\simeq 1.7$ for $L/\de=1000$, and $t/\tl\simeq 1.5$ for $L/\de=2000$.

In \figg{weibel}(c), we show that the Weibel phase is roughly independent of $\zeta$, with only marginal evidence for earlier Weibel onset times, higher linear growth rates, and greater saturation levels for larger $\zeta$. Yet, the saturation levels do not differ by more than a factor of two.

\figg{weibel}(d) shows that our results are well converged (apart from the case with 2 particles per cell per species in yellow, which cannot properly capture the exponential Weibel growth). The higher resolution case ($d_e=2$ cells, in blue) displays no significant difference in the growth rate or in the time evolution of the magnetic energy fraction as compared with our fiducial case, $d_e=1$. As discussed above, a resolution of $d_e=2$ or higher is required to capture the wavenumber of fastest Weibel growth with high accuracy. 

Finally, \figg{weibel}(e) shows that the magnetic energy fraction following the Weibel growth does not appreciably depend on the plasma temperature (or equivalently, on the turbulence Mach number).

\end{document}